\documentclass{optica-article}

\journal{opticajournal} 

\articletype{Research Article}

\usepackage{lineno}
\usepackage{subfig}

\begin{document}

\title{A systematic design approach for one-dimensional and crossed photonic nanobeam cavities for quantum dot integration}

\author{Oscar Camacho Ibarra,\authormark{1,*}
Jan-Gabriel Hartel,\authormark{1}
Atzin David Ruiz-Perez,\authormark{1}
Sonja Barkhofen,\authormark{1}
and Klaus D. J\"ons,\authormark{1}}

\address{\authormark{1}Institute for Photonic Quantum Systems (PhoQS),
Center for Optoelectronics and Photonics Paderborn (CeOPP) and
Department of Physics, Paderborn University,
Warburger Stra\ss e 100, 33098 Paderborn, Germany\\
\authormark{\dag}Oscar Camacho Ibarra and Jan-Gabriel Hartel contributed equally to this work.}

\email{\authormark{*}oscar.camacho.ibarra@uni-paderborn.de}

\begin{abstract*} 
We present a systematic workflow for the design of one-dimensional photonic crystal nanobeam cavities with non-zero cavity lengths. By simultaneously optimizing the lattice periodicity, air-hole geometry, and cavity length, our approach enables precise control of optical confinement while mitigating radiative losses and linewidth broadening effects. The method is further extended to the design of crossed nanobeam cavities with both matching and mismatched resonance frequencies. This strategy significantly reduces the need for extensive parameter sweeps, providing an efficient route toward optimized cavity designs for integrated quantum photonic applications. Moreover, the resulting structures are inherently compatible with the integration of single-photon emitters.

\end{abstract*}
\section{Introduction}
Embedding a multi-level quantum system within a cavity alters the way light and matter interacts. This field, known as cavity quantum electrodynamics, is both extensive and rich. In particular, when the interaction strength of the multi-level quantum system and the cavity mode is strong (the strong coupling regime SCR), the photon absorption and emission becomes a reversible process \cite{Brune1996,Dovzhenko2018}. The SCR is particularly relevant for quantum information processing \cite{Kimble2008} since at a single photon level it is possible to observe nonlinear optical effects \cite{Imamoḡlu1997} and prepare quantum states in a controlled and deterministic manner \cite{Haroche2013,Hofheinz2009}.

Nonlinear processes at the single-photon scale enable photon to photon interactions mediated by a single atom or an atomic ensemble \cite{Chang2014}. Such effects have being demonstrated using ensembles of Rydberg atoms \cite{Firstenberg2016} and through third-order nonlinearities in four-level atomic schemes \cite{Schmidt1996}. However, in both cases, to enable such interactions Electromagnetically Induced Transparency (EIT) was crucial in order to suppress absorption among certain energy levels.

To exploit nonlinear effects within cavity systems, EIT has also been explored in cavity quantum electrodynamics (cQED). Within the SCR in cQED, EIT and cavity-EIT are closely related phenomena. Standard EIT is a quantum interference effect where the presence of a control laser creates a narrow transparency window for a probe laser in an opaque medium \cite{Harris1997}.  In cavity EIT, this phenomenon is extended to systems where artificial atom(s) are coupled to an optical cavity. If the cavity and the atom are strongly coupled together then fine control of EIT is enabled \cite{Muecke2010}. This capability enables the photonic blockade effect \cite{Imamoḡlu1997,Birnbaum2005}, which is essential for the realization of quantum logic gates \cite{Werner1999}. Additionally, it supports the development of switches and routers critical for future quantum networks \cite{Brekenfeld2020}.

All the previously mentioned cEIT schemes require the employment of a system consisting of two cavities coupled to a single atom \cite{Heuck2020,Brekenfeld2020,Solak2024}, as well as high cooperativity, which in turn necessitates high quality factors ($Q$) and small mode volumes ($V$) \cite{Janitz2020}. Based on the seminal work in \cite{Majumdar2013}, we envisioned the implementation of a two cavity system through a photonic crystal (PC) platform for on-chip applications. In our case, a quantum dot (QD) will act as a quantum emitter, fulfilling the role of the atom in the previously described schemes. Since quantum dots (QDs) in InAs/GaAs can be monotonically integrated with a photonic crystal structure \cite{Sartison2022}, GaAs with air-holes is the  chosen platform for our PC cavities. QDs as integrated sources offer optical properties such as: ultra pure single-photon emission \cite{Somaschi2016}, high photon indistinguishability \cite{Ding2016}, on-demand generation of single photons \cite{Schweickert2018}, and emission spanning from the visible to the telecommunication wavelengths \cite{Paul2017}.
PC cavities offer advantages for the employment of a two cavity-atom coupled system, including scalability, stability, tunability and their lithographic fabrication allows for deterministic emitter placement with sub-20\,nm precision \cite{Janitz2020, Sartison2022}. Although atomic cavity QED systems may exhibit higher Q-factors, the ultra-small mode volumes of PC cavities yield comparable or even greater cooperativity \cite{Janitz2020}. Additionally, PC cavities can achieve near-unity coupling efficiencies into single-mode waveguides \cite{Arcari2014, Ding2016}, significantly outperforming atomic systems, which typically reach only around 34\% \cite{Ritter2012}.
Yet, finding the optimal parameters using state-of-the-art standard design methods to tailor single mode or bimodal nanobeam cavities for quantum dot integration is time consuming and often non-intuitive. Thus, in this work we present a methodical workflow for the design of 1D-photonic crystal nanobeam cavities and 1D crossed nanobeam cavities compatible with QD integration. In Sec.\ref{sec:II}, we briefly review previous works on the design of 1D nanobeam cavities. Sec.\ref{sec:III} introduces our two-parameter optimization scheme, which is then applied to a single nanobeam cavity in Sec.\ref{sec:IV}. In Sec.\ref{sec:V}, the method is extended to the crossed-cavity system, and finally, Sec.\ref{sec:VI} summarizes the resulting Q-factors and discusses the overall cavity performance. \vspace{3mm}

\section{Design Approaches}
Although the design and fabrication of one-dimensional photonic crystal nanobeam cavities has been extensively explored since the early 2000s, comparatively little attention has been given to tailoring these structures for the integration of epitaxial III–V semiconductor QDs. This gap arises because early designs were either not intended for quantum emitters/dot integration or did not account for the linewidth broadening of quantum dots near etched surfaces, an effect that was only reported in 2018 \cite{Liu2018}. Such broadening is detrimental, as it limits the photon indistinguishability required for realizing on-demand single-photon sources\cite{Liu2018}.

Most reported nanobeam cavity designs vary only a single structural parameter, such as the unit cell width \cite{Olthaus2019}, air-hole radius\cite{Quan2010,Quan2011}, or lattice periodicity \cite{Ohta2011}. By gradually tapering only one of these parameters, a local defect is introduced within the photonic bandgap, thereby forming a cavity while minimizing radiation losses to the environment. If the unit cell width is kept constant, the periodicity becomes the dominant parameter defining the effective dielectric filling fraction of the unit cell. To mitigate the linewidth broadening effects discussed in \cite{Liu2018} for the QDs, it is desirable to design cavities with a large dielectric fraction.

The exact periodicity range, where a bandgap opens, depends on the target wavelength, material system, and chosen unit cell geometry. However in practice, the periodicity values that yield the largest bandgap, thus high Q factors, generally fall within a range that does not satisfy the minimum separation of 150\,nm from an etched surface required to prevent linewidth broadening of QDs emission \cite{Liu2018}. Consequently, it becomes necessary to introduce a finite cavity length between the two central unit cells to provide sufficient spacing for quantum dot placement. Introducing a non-zero cavity length, however, leads to a new challenge: coupling between the localized Bloch mode supported by the photonic crystal and the waveguide mode of the central cavity region, which must be carefully addressed during cavity optimization.

To the best of our knowledge, only a few recent works \cite{Song2025,Vasco2020,Zhou2024} have proposed modified nanobeam cavity designs that explicitly consider the linewidth broadening effects reported in \cite{Liu2018}. However, these approaches do not constitute a generalizable design workflow, and the reported quality factors remain limited to the order of $10^{3}$. Alternatively, machine-learning-based optimization methods have been employed to design nanobeam cavities with quality factors on the order of $10^{4}$ \cite{Haagen2025,Liu2022}. Yet, not all of the resulting structures exhibit a large dielectric filling fraction of the unit cell or a sufficiently extended central cavity length for quantum dot integration.

Keeping these observations in mind, our workflow provides a deterministic approach for designing the cavity taper by simultaneously modifying two structural parameters: the periodicity and one geometric parameter defining the chosen air-hole shape. Our approach achieves a high quality factor of the order of $10^{5}$, and efficient coupling to the central region. Furthermore, our method can be extended to the design of more complex cavity coupled systems such as  crossed-cavity structures.

\label{sec:II}

\section{Map of Mirrior strenght}
\begin{figure}[htbp]
\centering\includegraphics[width=7cm]{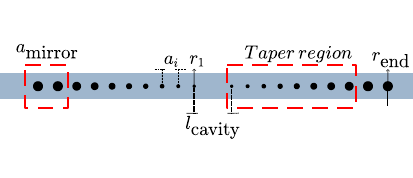}
\caption{Nanobeam cavity design: Symmetric 1D photonic crystal nanobeam cavity that incorporates a finite cavity length $l_{cavity}$, a mirror region and a taper region where the holes are tapered from small to big in order to maximize the coupling between the Bloch mode and the waveguide mode.}
\label{fig: 1DCavityDesign}
\end{figure}
\begin{figure}[htbp]
\centering\includegraphics[width=7cm]{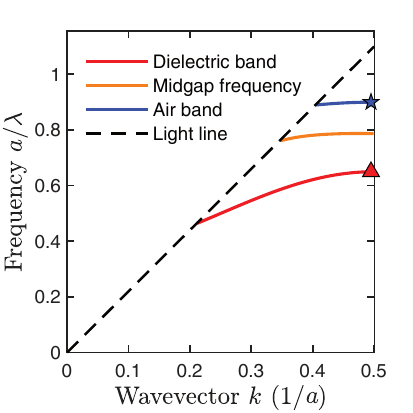}
\caption{Schematic illustration of a typical photonic bandstructure in the first half of the Brillouin zone.The dashed black line denotes the light line, separating radiative modes from modes confined to the structure; the dielectric (red) and air (blue) bands lie below it. The star and triangle mark the modes at the Brillouin zone edge, and the orange line indicates the midgap frequency that defines the center of the photonic bandgap.
}
\label{fig: Toymodel}
\end{figure}

The chosen nanobeam cavity design consists of three distinct regions: the cavity length, the taper region and the mirror region, as shown in Fig.\ref{fig: 1DCavityDesign}. By optimizing the parameters of the unit cells within each region, the light confinement in the cavity is tailored and a defect is introduced in the bandgap, thereby defining the cavity mode. In deterministic one-dimensional photonic crystal nanobeam cavities, this process is realized by modulating the reflectivity of the unit cells, with the mirror strength $\gamma$ model described in \cite{Quan2010}:

\begin{equation}
	\gamma = \sqrt{\frac{(\omega_{\text{air}}-{\omega_{\mathrm{di}}})^2}{(\omega_{\text{air}}+\omega_{\mathrm{di}})^2}-\frac{(\omega_{\text{target}}-\omega_{\mathrm{midgap}})^2}{(\omega_{\mathrm{midgap}})^2},}
	\label{eq:gammmaM}
\end{equation}
where $\omega_{\text{air}}$, $\omega_{\text{di}}$, $\omega_{\text{midgap}}$ and $\omega_{\text{target}}$ are the air band edge, dielectric band edge, mid-bandgap and target frequencies, respectively. Here, $\omega_{\text{target}}$ corresponds to the QD emission frequency, while $\omega_{\text{midgap}}$ is the midpoint between $\omega_{\text{air}}$ and $\omega_{\text{di}}$ (blue and red lines in Fig. \ref{fig: Toymodel}).

Besides the target frequency, the variables: $\omega_{\text{air}}$, $\omega_{\text{di}}$ and $\omega_{\text{midgap}}$ also vary depending on the unit cell parameters: periodicity $a$, hole size $r$ and the chosen hole shape. Consequently, to apply the model Eq.(\ref{eq:gammmaM}) it is necessary to compute the bandstructure of the unit cell for different parameters.

For a unit cell with a circular air hole, the design parameters are the periodicity $a$ and hole radius $r$. The width of the cavity, is fixed at 300\,nm corresponding with the width of a single mode rectangular waveguide in GaAs with a height of 163\,nm. In this context, the filling fraction $F$ is defined as the ratio of the air-hole area to the total unit cell area.

Our design approach starts with a total of 1571 single unit cell finite-difference time-domain (FDTD) simulations \cite{lum}. Through these simulations, the hole radius and the unit cell length are varied in order to obtain the air band and dielectric band values at the Brillouin zone edge ($k=0.5a$), indicated by the star and triangle respectively in Fig. \ref{fig: Toymodel}. The edge of the Brillouin zone is selected because it minimizes both the losses into the environment \cite{Quan2010} and the group velocity of the cavity mode. The resulting air-band and dielectric-band surfaces at the edge of the Brillouin zone are plotted in Fig.\ref{fig:3Dbands}, where the air band is represented by a blue color scale and the dielectric band by a red color scale. The target frequency of $321$THz corresponding to the QD emission is indicated by a gray plane, and its intersections with the dielectric and air bands are shown as black lines.

Using the values of $\omega_{\text{air}}$ and $\omega_{\text{di}}$ at the Brillouin edge, together with $\omega_{\text{midgap}}$ obtained from the simulations summarized in Fig. \ref{fig:3Dbands}, we applied the model in Eq.(\ref{eq:gammmaM}) for the target frequency of $321$THz. This yields the map of $\gamma$ shown in Fig. \ref{fig:mirrormap}. This map of $\gamma$ summarizes all combinations of unit cell parameters for which the target frequency lies within the photonic bandgap hence between the air band and dielectric band.

The advantage of using the map of $\gamma$ is that it unlocks the possibility of tuning the taper design by varying two parameters, in this case $a$ and $r$. While previous reported works achieved high Q-factors through the manipulation of single parameter for the tapering region, either $a$ \cite{Ohta2011}  or $r$ \cite{Quan2010,Quan2011}. In contrast, a simultaneous manipulation of $a$ and $r$ enables a fine control of the location of the targeted frequency inside the bandgap. The designs reported in \cite{Ohta2011} and \cite{Quan2010,Quan2011} would qualitatively correspond to the solid green and orange straight lines in our $\gamma$ map (Fig.~\ref{fig:mirrormap}), noting that a direct comparison is not possible because of differences in the chosen material platform between works. Nevertheless, we recall that without simultaneously varying both parameters, it is not possible to realize a taper that starts from a small central hole on the dielectric band and progresses toward a higher $\gamma$. Moreover, by summarizing all possible tapering strategies, the map of $\gamma$ offers increased flexibility in parameter selection and facilitates adaptation in the presence of fabrication constraints. In particular, it enables the construction of arbitrary taper profiles directly from the map, as discussed below. \vspace{3mm} 
\label{sec:III}
\section{Design method 1D nanobeam cavity}
\begin{figure}[htbp]
\centering\includegraphics[width=14cm]{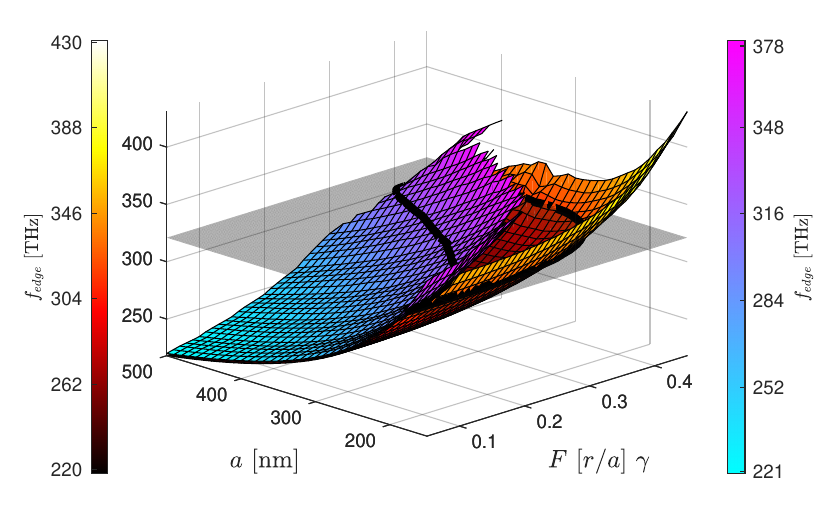}
\caption{ Variations of the (TE) dielectric and air bands as a function of the lattice constant ($a$) and filling fraction ($F$). The warm colormap represents the dielectric band, while the cool colormap represents the air band. The gray plane at $321$THz indicates the bandgap region, with intersections with the dielectric and air bands shown as black lines.}.
\label{fig:3Dbands}
\end{figure}

We design our cavity to be resonant with the most commonly used In(Ga,As)/GaAs quantum dots \cite{Wijitpatima2024, Goldstein1985, Jin2005} emitting at 934\,nm ($321$THz). Our material platform of choice is GaAs, which supports monolithic integration of these quantum dots embedded in the same matrix via marker-based localization approach \cite{Madigawa2023, Peniakov2024, Copeland2024}, and offers a high refractive index contrast with air, enabling strong optical confinement in suspended nanobeam cavities.

For our cavity design two parameters of the cavity are fixed: The first is the cavity height, which is set to 163\,nm. This value corresponds to the thickness of the GaAs layer, which is determined by the epitaxial growth process and optimized for strong optical confinement of the fundamental mode. The second, the cavity width, that is constrained to 300\,nm which corresponds to the width of a single mode waveguide with a rectangular cross-section in GaAs and height of 163\,nm. The remaining design parameters to optimize are: the unit cell length ($a$), the cavity length ($l$) and the air-hole radii ($r$) in both the mirror region and the tapering region.

\begin{figure}[htbp]
	\centering
	
	\centering\includegraphics[width=7cm]{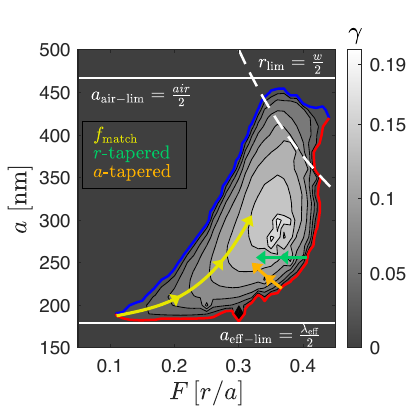}
	\caption{ Map of $\gamma$ at the target frequency of $321$\,THz, obtained using the model of $\gamma$ \cite{Quan2010}\cite{Quan2011} over the bandgap area of Fig.\ref{fig:3Dbands}. The color bar indicates the value of $\gamma$ across the parameter map. The dashed white lines denote the threshold at which the hole size exceeds the waveguide width. The green and orange lines illustrate how design strategies based solely on the lattice periodicity \cite{Ohta2011} or the air-hole geometry \cite{Quan2010,Quan2011} would appear on the $\gamma$ map. Red segments of the outer contour indicate intersections with the dielectric band, while blue segments indicate intersections with the air band.}

	\label{fig:mirrormap}
\end{figure}

In our workflow, the optimization of the cavity length ($l$) is postponed until the end, while the remaining parameters are addressed first. Specifically, the unit cell length ($a$) and the hole radii ($r$) of both the mirror and the tapering region are extracted from the map of $\gamma$ ( Fig.\ref{fig:mirrormap}) for a target frequency (wavelength).

Recall that a cavity is formed by having a defect mode inside the band-gap. Thus, the tapering of the periodicity and holes fulfills 3 key requirements for the cavity: the first unit cell adjacent to the cavity length $l_{cavity}$, defined by $a_1$ and $r_1$, has its dielectric band aligned with the target frequency, second it gradually modulates the reflectivity $\gamma$ to minimize radiative losses into the environment \cite{Akahane2003} and third, by employing a taper in which the central holes are smaller than those in the mirror region, the coupling between the Bloch mode and the waveguide mode is enhanced \cite{Wang2022}. The location of the cavity mode maximum depends on the chosen band for the defect, either dielectric or air. Although the air band (blue line in Fig.\ref{fig:3Dbands}) can also be used as the defect mode \cite{Quan2011}, the requirement to place a QD restricts our design to the dielectric band (red line Fig.\ref{fig:mirrormap}).

We now examine the map of $\gamma$ shown in Fig.\ref{fig:mirrormap}, it will tell us how to build the system. The axes correspond to the periodicity $a$ and the filling fraction $F = r/a$, while the color scale indicates the reflectivity $\gamma$. Dashed white lines mark the limits beyond which the hole size exceeds the waveguide width, and is thus not implementable for free standing photonic structures. The map of $\gamma$ is formed by multiple contour lines, each representing a constant value of $\gamma$. The lower and right portions (red line) of the outer contour line corresponds to the intersection between the resonance frequency and the dielectric band. While the upper and left portions (blue line) of the outer contour line corresponds to the intersection between the resonance frequency and the air band. Hence, tracing a path from the red edge toward the center with high reflectivity alters the values of the periodicity ($a$) and the hole radii ($r$), thereby defining the parameters for each unit cell of the taper. In other words, the choice of path defines the radius size on each unit cell and hence the taper parameters. In our design the taper region contains 15 unit cells, while the mirror region consists of 5 unit cells. A convergence study of the $Q$ factor with respect to the number of taper unit cells shows saturation at 15. The additional 5 mirror unit cells increase the reflectivity of the structure while keeping the cavity compact. 
 
For the cavity design $f_{match}$ presented here, the chosen path is highlighted in yellow in Fig.\ref{fig:mirrormap}. Our choice maximizes the coupling to the waveguide at the cavity center by placing the smaller radius holes at the center. As demonstrated in \cite{Wang2022}, near-unity coupling efficiency into a feeding-waveguide is possible when the target frequency $\omega_{\text{target}}$ matches with the middle of the bandgap. Furthermore, this selection also maximizes the $\gamma$ as can be seen from Fig.\ref{fig:mirrormap}.

\begin{figure}[htbp]
    \centering

    \captionsetup[subfloat]{position=top, justification=raggedright,
        labelformat=parens, labelsep=space}

    \subfloat[]{%
        \includegraphics[width=0.465\columnwidth]{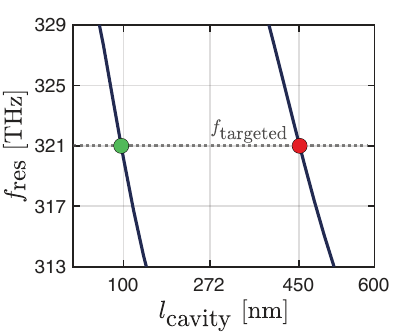}
        \label{fig:cavityLength}
    }\hfill
    \subfloat[]{%
        \includegraphics[width=0.48\columnwidth]{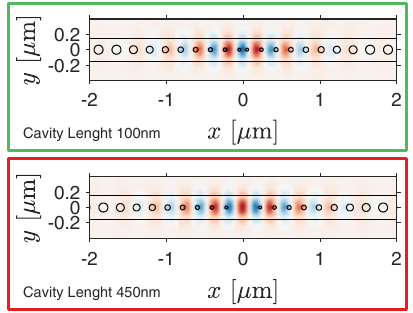}
        \label{fig:FieldPBC}
    }

    \caption{(a) Cavity resonance frequency versus cavity length $l$, obtained by extending the dielectric region at the cavity center between the two central holes and sweeping its length from $0$\,nm to $600$\,nm. The targeted frequency $f_\text{targeted}=321$,THz is indicated by the grey dashed line. Resonances at the target frequency occur for cavity lengths of approximately $100$\,nm (green dot) and $450$\,nm (red dot). (b) $E_y$-field profile for a cavity length of $100$\,nm and $450$\,nm. Black lines denote the structure outlines of the waveguide and the holes. }
    
\end{figure}

The only remaining parameter to be optimized, in order to match the target frequency at $321$\,THz, is the cavity length $l$. To adjust $l$, the dielectric region at the cavity center between the two central holes is extended, and its value is swept from $0$\,nm to $600$\,nm while simultaneously tracking the cavity resonance (Fig.\ref{fig:cavityLength}). Our simulations show resonances at the target frequency for cavity lengths of $100$\,nm and $450$\,nm.

Considering the size of a quantum dot (QD) to be approximately $50$\,nm \cite{Verma2022}, and considering a placement error of $\pm30$\,nm relative to the cavity center \cite{Madigawa2023}, larger cavities are easier to fabricate with high yield. Furthermore to prevent any spectral broadening of the QD emission caused by nearby air boundaries \cite{Liu2018}, we select $l=450$\,nm (see Fig.\ref{fig:FieldPBC}). While further increasing $l$ could reduce the risk of spectral broadening, it would also lead to an undesirable increase in mode volume.

The design steps can be summarized as follows:
\begin{enumerate}
	\item Adjust the width ($w$) of the nanobeam cavity to match that of a single-mode ridge waveguide of identical height at the targeted frequency.\vspace{1mm} 
	
	\item Compute the bandstructure over a broad parameter set of values for $a$ and $r$, and extract the dielectric and air band values at the Brillouin zone edge. Here, $n_{\text{eff}}$ denotes the effective refractive index of the guided mode at the design wavelength and the condition:
	\begin{equation}
		a = \frac{\lambda}{2 n_{\text{eff}}}
		\label{eq:neff}
	\end{equation}
	is used to compute the upper and lower bounds for $a$. While the upper bound for $r$ will be given by $w = 2r$. This upper bound corresponds to the case where the air holes are sufficiently large to fragment the nanobeam cavity into isolated dielectric segments. \vspace{1mm} 
	
	\item Compute the map of $\gamma$ using the model in Eq.(\ref{eq:gammmaM}) and the chosen target frequency.\vspace{1mm}

	\item As mentioned previously, the bottom outer contour red line on the map of $\gamma$ corresponds to the case where $\omega_{\text{target}} =\omega_{\text{di}}$. Any point on this bottom lane will define the parameters for the central unit cell ($a_1$ and $r_1$), i.e., this will be the defect mode of the cavity.\vspace{1mm}
	
	\item Multiple paths can be defined from the bottom lane. The path that maximizes the reflectivity satisfies the condition $\omega_{\text{target}} = \omega_{\text{midgap}}$.\vspace{1mm}
	
	\item Introduce the cavity length $l$ and sweep its value until the resonance at the targeted frequency is obtained. 
\end{enumerate}
The calculated $Q$ and $V$ values for our design are summarized in the first line of Tab.\ref{tab:QandV}.

\label{sec:IV}
\section{Design method for Cross-cavity}

\begin{figure}[htbp]
    \centering

    \captionsetup[subfloat]{position=top, justification=raggedright,
        labelformat=parens, labelsep=space}

    \subfloat[]{%
        \includegraphics[width=0.465\columnwidth]{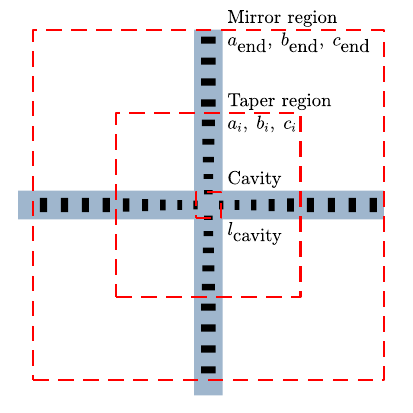}
        \label{fig:CrossSchema}
    }\hfill
    \subfloat[]{%
        \includegraphics[width=0.48\columnwidth]{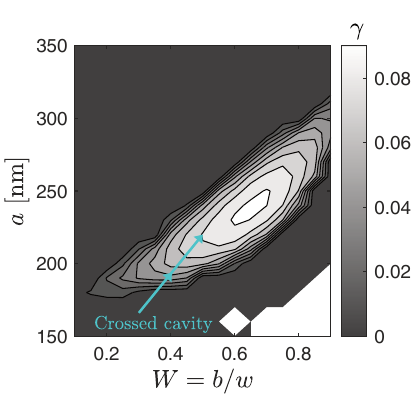}
        \label{fig:GammaRec}
    }

    \caption{(a) Schematic of the crossed cavity design, where each cavity has the same frequency. (b) Map of $\gamma$ for a rectangular hole shape at $\omega_{\text{target}}=321$THz. The turquoise line shows the chosen arbitrary path.}
    \label{fig:crosscavity}
\end{figure}

Based on our map of $\gamma$ design method, we now extend our approach to realize crossed 1D photonic crystal cavities, as schematically illustrated in Fig.\ref{fig:CrossSchema}. These structures have been sparsely investigated in previous works. To the best of our knowledge, the only prior works are found in \cite{Rivoire2011,Rivoire2011a} where they report 1D-crossed nanobeam cavities with Q factors up to $1.9\times10^{4}$. Additionally, a similar work is presented in \cite{Fehler2019} where a nanobeam cavity system is intersected orthogonally by a probe waveguide. Lastly in \cite{Johnson1998}, two orthogonal 2D photonic crystal waveguides with minimum cross-talk are reported.

The computed map of $\gamma$ for the nanobeam cavity with rectangular holes is shown in Fig.\ref{fig:GammaRec}. To illustrate its effectiveness, an arbitrary path (turquoise line) was selected. Unlike the 1D cavity approach, this path does not begin at the lower contour of the $\gamma$ map but instead starts five unit cells earlier within the region of zero $\gamma$. This choice accounts for the strong influence of the crossed arm on the unit cells near the cavity center. As observed in the simulations, the unit cells located closer to the center will exhibit no resonance modes. Although the unit cells closer to the center do not exhibit resonance modes, they still play a crucial role in facilitating a smoother transition between the Bloch mode and the waveguide mode. The simulations further show that the location of the defect mode at $321$\,THz shifts to the third unit cell for the case of a cross-cavity system with matching frequencies on each cavity arm. In addition, the perturbation introduced by the cavity crossing leads to a small frequency shift, such that the resonance appears at approximately $320.2$\,THz instead of the targeted $321$\,THz.

\begin{figure}[htbp]
	\centering
	\centering\includegraphics[width=7cm]{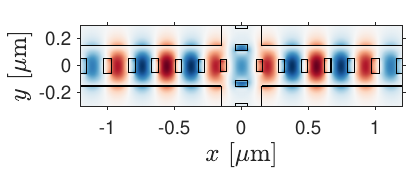}
	\caption{ $E_y$ field distribution for a crossed cavity with matching frequencies at $\omega_{\text{target}}=320.2$THz. Black lines represent structure outlines. At the origin, the field strength is $0.5717\cdot E_{max}$.}
	\label{fig:FieldCrossCav}
\end{figure}

Thus, our workflow for designing 1D nanobeam cavities can be extended to the cross-cavity system with minor modifications:

\begin{itemize}
	\item Rectangular air holes are employed instead of circular ones. Simulations show that the crossed-cavity design yields $Q$ factors on the order of $10^3$ when implemented with circular air holes, whereas with rectangular air holes the $Q$ increases to the order of $10^5$.
	
	\item The unit cell dimensions are parameterized as illustrated in Fig.\ref{fig:CrossSchema}, where $a$ represents the periodicity, $w$ the cavity width, and $b$ and $c$ are the height and width of the rectangular air-hole, respectively. The parameter $c$ is fixed to $c=b/2$ (this choice was arbitrary), therefore only $b$ will be swept while obtaining $\omega_{\text{air}}$ and $\omega_{\text{di}}$.
	
	\item The crossed section at the center is not treated as a PC with a different unit cell, but rather as a perturbation of the PC which introduces scattering losses. Because the crossing suppresses the resonances of the central unit cells, the defect mode at the target frequency is instead supported by the adjacent unit cells. Consequently, the path is designed to start five unit cells earlier within the zero-reflectivity region, see Fig.\ref{fig:GammaRec}.
\end{itemize}

Alternative cavity-QD schemes like \cite{Richter_2025} require multiple off-resonant pulses. Therefore, it is also of interest to design a cross-cavity system with non-matching frequencies between the cavity arms. For our workflow this would simply imply computing a second map of $\gamma$ for the second frequency. In principle, the detuning can be along any frequency inside the bandgap, hence between the air and dielectric planes in Fig.\ref{fig:3Dbands}.

\begin{figure}[htbp]
    \centering
    \captionsetup[subfloat]{position=top, justification=raggedright,
        labelformat=parens, labelsep=space}

    \subfloat[]{%
        \includegraphics[width=0.48\columnwidth]{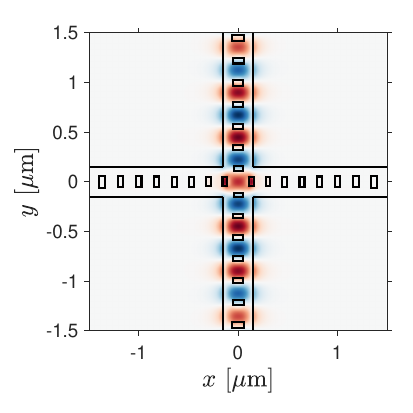}
        \label{fig:AssyEFields322}
    }\hfill
    \subfloat[]{%
        \includegraphics[width=0.48\columnwidth]{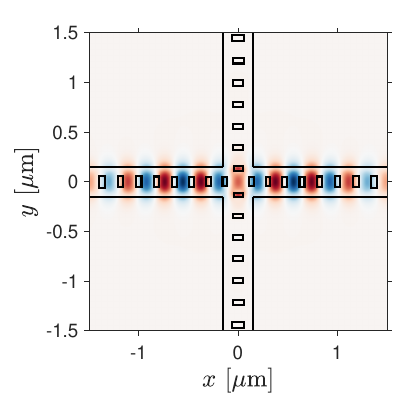}
        \label{fig:AssyEFields289}
    }

    \caption{Unmatched-$f$ cavity design. (a) $E_y$ field profile for the $x$-arm at $322.9$,THz; (b) $E_x$ field profile for the $y$-arm at $289.1$\,THz.}
    \label{fig:crosscavity_unmatched}
\end{figure}

As an example, we report on a design in which the second cavity (the x-arm) is redshifted by $100$\,nm relative to the target wavelength of $934$\,nm ($321$\,THz) used for the y-arm, although a blueshift could likewise be employed. The resulting $E$ field profiles are shown on Fig.\ref{fig:AssyEFields322} and Fig.\ref{fig:AssyEFields289}, and similarly to the matching frequency design, the location of the defect modes at $321$\,THz and $289$\,THz will shift to the third unit cell for the case of a cross-cavity system with matchless frequencies on each cavity arm. Due to the perturbation introduced by the cavity crossing, the resonance designed at $321$\,THz appears at $322.9$\,THz in this configuration, while the second cavity mode occurs at $289.1$\,THz. Table \ref{tab:QandV}, lines 2-4 , summarizes the Q and V for the unmatching frequency design and the previous designs.

\label{sec:V}
\section{Conclusion}
\begin{table*}[ht]
	\centering
	\caption{Q and V values for the different Cavities designs}
	\begin{tabular}{lccc}
		\hline
		\textbf{Cavity} & \textbf{Frequency (THz)} & \textbf{Quality Factor ($Q$)} & \textbf{Mode Volume ($V$)} \\
		\hline
		1D Cavity         & 321   & $5.6 \times 10^5$ & $0.80 \lambda^3 / n^3$ \\
		Cross-cavity Matched $f$      & 321   & $4.6 \times 10^5$ & $1.27 \lambda^3 / n^3$ \\
		Cross-cavity Unmatched $f$ $y_{\text{arm}}$   & 322.9 & $5.2 \times 10^4$ & $1.22 \lambda^3 / n^3$ \\
		Cross-cavity Unmatched $f$ $x_{\text{arm}}$  & 289.1 & $6.4 \times 10^4$ & $1.30 \lambda^3 / n^3$ \\
		\hline
	\end{tabular}
	\label{tab:QandV}
\end{table*}

In this work, we have presented a workflow for designing one-dimensional photonic crystal nanobeam cavities with non-zero cavity lengths by optimizing three key parameters: the lattice periodicity, a characteristic geometric parameter of the chosen air-hole shape, and the central cavity length. By simultaneously manipulating the first two parameters, the periodicity and the air-hole geometry, we are able to incorporate features previously demonstrated in separate works, namely enhanced coupling between the Bloch mode and the waveguide mode \cite{Wang2022}, and reduced radiative losses into the environment \cite{Akahane2003}. The introduction of the third parameter, the cavity length, further allows us to mitigate linewidth broadening effects as discussed in \cite{Liu2018}. Furthermore, we extend this design methodology to the deterministic realization of crossed nanobeam cavities, enabling both matching and mismatched frequency configurations. Overall, our approach reduces the need for extensive parameter sweeps by providing a more efficient and systematic route for cavity designs.

Future research on 1D PC crossed-cavity systems may build upon the framework presented here by performing a more detailed analysis of the central cavity intersection, with the aim of further reducing optical losses and enhancing overall device performance. Alternative approaches, such as utilizing inverse design techniques based on topological optimization, may also enable the realization of low-loss central intersections. Furthermore, implementing crossed cavity designs based on two distinct topological modes of 1D photonic crystals represent another promising avenue for reducing the losses at the central intersection and reducing the mode volumes. \vspace{3mm} 

\label{sec:VI}

\section{Acknowledgments}
We gratefully acknowledge Samuel Gyger for helpful discussions. This work is supported by the Deutsche Forschungsgemeinschaft (German Research Foundation) through the transregional collaborative research center TRR142/3-2022 and by the European Research Council (ERC) under the European Union’s Horizon 2020 research and innovation program (LiNQs, 101042672). K.D.J. acknowledges funding from the Ministry of Culture and Science of North Rhine-Westphalia for the Institute for Photonic Quantum Systems (PhoQS).
\label{sec:VII}
\bibliography{references}

\end{document}